\documentclass[showpacs,preprintnumbers,amsmath,amssymb]{revtex4}

\usepackage{graphicx}
\usepackage{dcolumn}
\usepackage{bm}
\newcommand{\me}{\mathrm{e}}%
\newcommand{\mi}{\mathrm{i}}%
\newcommand{\dif}{\mathrm{d}}%
\begin{document}

\title{Instanton Effects in QCD Sum Rules for the $0^{++}$ Hybrid}

\author{Zhu-feng Zhang$^1$, Qing Xu$^2$, Hong-ying Jin$^2$ and T. G.  Steele$^3$\\
$^1$Physics Department, Ningbo University, Zhejiang Province, P. R. China\\
$^2$Zhejiang Institute of Modern Physics, Zhejiang University, Zhejiang Province, P. R. China\\
$^3$ Department of Physics and Engineering Physics,
University of Saskatchewan, Saskatoon, Saskatchewan, Canada S7N 5E2 }

\date{\today}

\begin{abstract}
In this paper, we study instanton contributions to the correlator of the hybrid current
$g\bar q \sigma_{\mu\nu}G^a_{\nu\mu}T^a q$.
These contributions are then included in a QCD sum-rule analysis
of the  isoscalar $0^{++}$ hybrid mass.
We find a mass at $1.83\,\textrm{GeV}$ for the $(\bar uug+\bar ddg)/\sqrt{2}$
hybrid. However, for the $\bar ssg$ hybrid, we find the sum rules are unstable. We also study  non-zero width effects, which affect
the mass prediction. The mixing effects between these two states are studied and
we find  QCD sum rules support the existence of a flavor singlet hybrid with mass at around 1.9\,GeV. Finally,
we study the mixing effects between hybrid and glueball currents.  The mixing between the $(\bar uug+\bar ddg)/\sqrt{2}$($\bar ssg$) and the glueball causes two states, one in the region 1.4-1.8\,GeV(1.4-2.2\,GeV), and the other  in the range  1.8-2.2\,GeV(2.2-2.6\,GeV).
\end{abstract}

\maketitle

\section{Introduction}

It is generally believed that hadrons beyond the conventional quark model may exist.
Theoretical predications of hybrid properties (mass, decay constant etc.) are therefore an important ingredient of their experimental confirmation.

Among these unconventional hadrons, the hybrids (first predicted in \cite{Chodos:1974pn,Jaffe:1975fd}) have attracted considerable attention.
Hybrid properties have been studied in many approaches, such as  the Bag Model\cite{Hasenfratz:1980jv,Barnes:1982tx}, Potential
Models\cite{Horn:1977rq} and QCD sum rules\cite{Balitsky:1982ps,Balitsky:1986hf,Narison:1999hg} (see  \cite{Huang:2011nv,Chen:2010ic,Huang:2010dc,Qiao:2010zh} for the most recent works).   Most studies have focused on  hybrids with exotic quantum numbers (e.g.,  $1^{-+}$) because such states do not mix with ordinary $\bar qq$ mesons.
Less attention has been paid to the $0^{++}$ hybrid because  it is experimentally very difficult to identify the composition of such a state,  and it is also considered as a highly excited flux tube (e.g., some authors think the $0^{++}$ is one of the degenerate ground states in the flux tube model\cite{recentfluxtube}).

Huang et al previously studied the hybrid with $0^{++}$  quantum numbers in the framework of QCD sum rules\cite{Huang:1998zj}. They used two currents: $g\bar q\sigma_{\mu\nu}G^a_{\nu\mu}T^aq$ and $g\bar q\gamma_\mu G^a_{\nu\mu}T^aq$. They found that
the scalar current predicts $\bar ssg$ hybrid at a mass of $2.30-2.35\,\textrm{GeV}$ while the vector current predicts $3.4\,\textrm{GeV}$.  The interesting thing is that the predicted $0^{++}$ hybrid mass ($2.30-2.35$GeV) is much lower than the previous predictions\cite{Isgur:1985vy,Merlin:1986tz,Govaerts:1983ka,Govaerts:1984bk,Govaerts:1986pp}. This $2.30\,{\rm GeV}$ prediction may be meaningful in experiments. From the particle data book, we find there are four $0^{++}$ mesons in the region $2000\,{\rm MeV}\sim 2300\,{\rm MeV}$. This is more crowded than the group of $f_0(1370)$, $f_0(1500)$ and $f_0(1710)$. Therefore, even if the glueball state is included, it is still hard to explain these four states. One possible explanation may be that there is a  $0^{++}$ hybrid among them. But this picture is not compatible with the flux tube picture, so it is worth studying the $0^{++}$ hybrid mass in more detail.

The scalar  channels are known to contain  important effects from instantons\cite{'tHooft:1976fv,Schafer:1996wv}, which play an important role in the scalar glueball and pseudoscalar glueball calculations. Thus instantons may also be important for the $0^{++}$ hybrid. In this paper, we calculate such instanton effects to the correlator of the $0^{++}$ hybrid current $g\bar q \sigma_{\mu\nu}
G^a_{\nu\mu} T^a q$. With our new result supplementing other known  contributions, we calculate the hybrid mass with quark content $u,d$ and $s$ respectively.
In addition, we study the effect of a Breit-Wigner form for the phenomenological spectral density.
Flavor mixing effects between these two states are then studied in an attempt to give a prediction of the mass of flavor octet
and singlet scalar hybrid. Finally, we study the mixing effects between hybrid and glueball currents.

\section{The masses of the pure states: $(\bar uug+\bar ddg)/\sqrt 2$ and $\bar ssg$}

We start our QCD sum rule analysis by considering two scalar hybrid currents as used in Ref.\cite{Huang:1998zj} which are pure states in flavor space with isospin 0, i.e.,
\begin{equation}
j_1(x)=\frac{1}{\sqrt 2} g[\bar u(x) \sigma_{\mu\nu} G^a_{\nu\mu}(x)T^a u(x)+\bar d(x)\sigma_{\mu\nu} G^a_{\nu\mu}(x) T^a d(x)],
\end{equation}
and
\begin{equation}
j_2(x)= g\bar s(x) \sigma_{\mu\nu} G^a_{\nu\mu}(x)T^a s(x).
\end{equation}

The first step of  QCD sum rules is to  use the operator-product
expansion (OPE) to calculate the correlation function, which is defined as \cite{Shifman:1978bx,Shifman:1978by,Novikov:1983gd}:
\begin{equation}
\Pi(q^2)=\mi\int\dif^4 x\, \me^{\mi q x}\langle 0|Tj(x)j^\dagger(0)|0\rangle.
\end{equation}

The imaginary part of the correlator (up to dimension 6 condensates) is given in   \cite{Huang:1998zj}:
\begin{equation}
\label{eq:impi0}
{\rm Im}\Pi(s)^{\textrm{(OPE)}}=\frac{\alpha_s(\mu^2)}{24\pi^2} s^3+4\alpha_s(\mu^2) m_q \langle\bar qq\rangle s-m_q^2
\langle \alpha_s G^2\rangle+\frac{8\alpha_s(\mu^2)\pi^2}{3}m_q^2\langle \bar qq\rangle^2\delta(s),
\end{equation}
where $q=u$ for $j_1$ and $q=s$ for $j_2$.

The second standard step of QCD sum rules is to combine the OPE-correlator with the phenomenological single narrow resonance spectral density ansatz
\begin{equation}
\label{eq:single}
{\rm Im}\Pi(s)^{\textrm{(phen)}}=\pi f^2_H
(m^2_H)^3 \delta(s-m_H^2)+ {\rm Im}\Pi(s)^{{\rm (OPE)}}\theta(s-s_0)
\end{equation}
via the Borel-transformed correlator
\begin{equation}
\hat B \Pi(q^2)\equiv \lim_{\substack{Q^2,n\rightarrow \infty\\ n/Q^2=\tau}}
\frac{(Q^2)^{n+1}}{n!}\left(- \frac{\dif}{\dif Q^2}\right)^n\Pi(-Q^2)
=\frac{1}{\pi}\int_0^\infty \dif s \,\me^{-s\tau} {\rm Im }\Pi(s),
\end{equation}
then we reach the sum rule:
\begin{equation}
\label{eq:newsr}
f_H^2 (m_H^2)^3 \me^{-m_H^2 \tau}+\frac{1}{\pi}\int_{s_0}^\infty\dif s\, \me^{-s\tau}{\rm Im}\Pi(s)^{{\rm (OPE)}}
=\hat B\Pi(q^2)^{{\rm (OPE)}},
\end{equation}
where $s_0$ is the threshold separating the contribution
from higher excited states and the QCD continuum.

Notice that in Eq.\eqref{eq:impi0}, the leading-order mass corrections for the $\langle \alpha_s G^2\rangle$ term
and $\langle \bar qq\rangle^2$ term, which should play an important role in the OPE of correlator,
 are not present (i.e., there is a chiral suppression of these terms). Because of the lack of these terms, the non-perturbative contributions may be underestimated.
This defect may be improved by introducing instanton effects into the sum rules.

It has been known for a long time that the instanton plays an important role in the QCD vacuum and hadron
physics\cite{'tHooft:1976fv,Schafer:1996wv}. The explicit instanton in regular gauge can be expressed as
\begin{equation}
 A_\mu^a(x;z) =\frac{2}{g} \frac{ \eta_{a\mu\nu} (x-z)_\nu}{(x-z)^2+\rho^2},
\end{equation}
where $\rho$ is the instanton size, $z$ is the position of the instanton center in Euclidean space,
and $\eta$ is the 't Hooft symbol.

If an instanton exists in space, quarks may occupy a special state called zero-mode state (one of each flavor).
Usually, the quark zero-mode propagator is complicated, but if we deal with problems
in the framework of Single-Instanton Approximation (SIA)\cite{Faccioli:2001ug}, the propagator takes a simple form,
\begin{equation}
S(x,y;z) =  \frac{\rho^2}{8 \pi^2 m^\star}
\frac{1} { ((x-z)^2+\rho^2)^{3/2} ((y-z)^2+\rho^2)^{3/2}}
\left[\gamma^\mu \gamma^\nu\frac{1}{2} (1-\gamma^5)\right]
\otimes (\tau^+_\mu \tau^-_\nu),
\end{equation}
where $\tau^\pm =(\bm{\tau},\mp\mi)$, and $m^\star$ is the quark effective mass which takes all the collective
contribution of all instantons other than the leading one. For $u, d$ quarks, after evaluating the effective mass
in the  Random Instanton Liquid Model and Interacting Instanton Liquid Model, Faccioli et al pointed out that
$m_u^\star=86\,\textrm{MeV}$ should be used in the applications of SIA when two zero-mode propagators are involved
\cite{Faccioli:2001ug}. However, in the present case, besides quark zero-modes, we also have direct instanton
contributions from gauge field strength. This extra contribution may change the value of $m^\star$ away
from 86\,MeV. Thus, in this paper, we will still use the formerly widely used estimate by Shifman et al\cite{Shifman:1979uw}:
\begin{equation}
m_q^\star=m_q-\frac{2}{3}\pi^2 \rho^2 \langle \bar qq\rangle.
\end{equation}

\begin{figure}[htbp]
\centering
\includegraphics[scale=0.6]{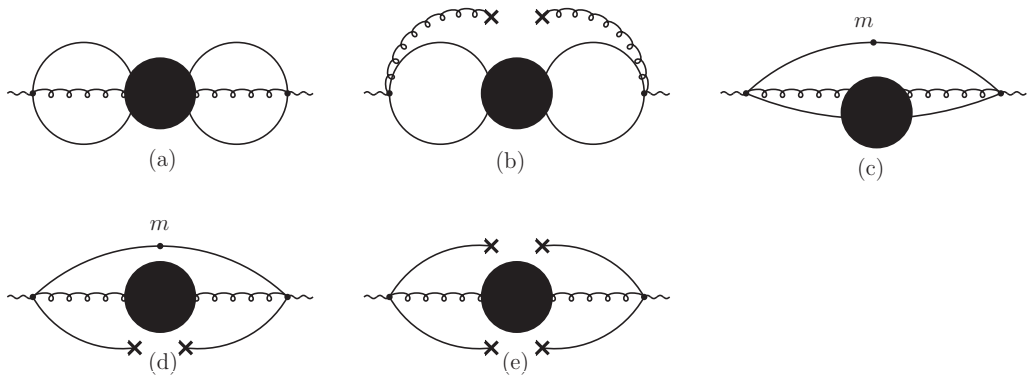}
\caption{\label{fig:inst} Non-zero instanton contributions for the correlator of $j_1$ (Fig.(a-e)) or $j_2$ (Fig.(c-e)). (The
blob denotes an instanton.)}
\end{figure}

\begin{figure}[htbp]
\centering
\includegraphics[scale=0.6]{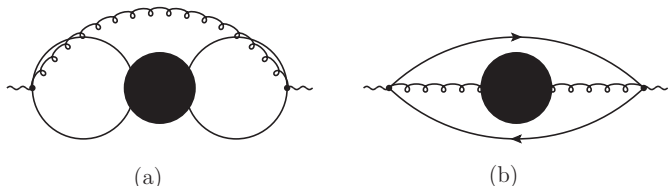}
\caption{\label{fig:zeroinst} Some zero instanton contributions.}
\end{figure}

Now let us consider the instanton contributions to the correlation functions of currents
$j_1$ and $j_2$. In Fig.\ref{fig:inst} we draw some diagrams which are associated with nonvanishing contributions.
The instanton contributions originating from Fig.\ref{fig:inst}(a-b) only play a role in the correlator of $j_1$,
while Fig.\ref{fig:inst}(c-e) contribute to both $j_1$ and $j_2$. There are many other figures with  vanishing contributions to the correlator (e.g. Fig.\ref{fig:zeroinst}) that explains why the leading-order mass corrections to the $\langle \bar qq\rangle^2$
and $\langle \alpha_s G^2\rangle$ condensates in Eq.~\eqref{eq:impi0} do not appear.

Among these instanton contributions, the most important one for $j_1$ comes from Fig.\ref{fig:inst}(a), which reads
\begin{equation}
\label{eq:inst0}
\Pi(q^2)_{j_1}^1=\frac{1}{2}\int \dif \rho \frac{n(\rho)\rho^2}{m_u^{\star 2}} Q^6 K^2_3(Q\rho),
\end{equation}
where $n(\rho)$ is the density of instantons. We get the imaginary part of $\Pi(q^2)^1_{j_1}$ upon analytic continuation to
the physical domain,
\begin{equation}
\label{eq:inst00}
{\rm Im}\Pi(s)_{j_1}^1=-\int \dif \rho \frac{n(\rho)\rho^2\pi^2}{4 m_u^{\star 2}} s^3 J_3(\sqrt{s}\rho)Y_3(\sqrt{s}\rho).
\end{equation}

Usually one uses the simple spike distribution $n(\rho)=\bar n\delta(\rho-\bar\rho)$, where
$\bar n =8\times 10^{-4}\,\textrm{GeV}^4$ is the average instanton density and $\bar \rho=1/(0.6\,{\rm GeV})$
is the average instanton size. Notice that the instanton contributions (Eq.\eqref{eq:inst00} and that from the other
three diagrams) are oscillating functions, of which  the amplitude  becomes larger along with s in some regions where
the spectral density is negative. This problem also arises in other cases, for instance see \cite{Forkel:2003mk,Zhang:2003mr}.
This situation is called local duality violation \cite{shifman:2000}.
One approach used to mitigate this unphysical tendency introduces  a distribution function of the instanton, for
instance, in Ref.\cite{Forkel:2003mk}, the author used a gaussian-tail distribution
\begin{equation}
n(\rho)=\frac{2^{18}}{3^6\pi^3}\frac{\bar n}{\bar \rho}\left(\frac{\rho}{\bar \rho}\right)^4 \me^{-2^6\rho^2/
(3^2\pi \bar \rho^2)}
\end{equation}
in the analysis of QCD sum rules for glueball, which improves the physical tendency of the spectral density.
However, this approach of a modified instanton distribution does not seem to be intrinsically necessary. For example, local dulaity violation has also been addressed by constructing differently-weighted sum-rules \cite{Zhang:2003mr}.
For the cases we are studying, the gaussian tail distribution does not work.

In our case, the amplitude of the instanton-induced spectral density  does not decline quickly when $s $ becomes larger; this means the contributions from large instantons are not suppressed, and thus the single instanton approximation is not
appropriate in the Minkowski domain. However, in the deep Euclidean domain, instanton contributions are exponentially suppressed, and  the single instanton approximation works well. That means global duality is not violated  although
local duality is strongly violated in the single instanton approximation. Because  QCD sum rules are based global duality,  the single instanton approximation should be  appropriate.

Based on the considerations above, we deal with instanton contributions in the  Euclidean domain
 in this paper, i.e., we introduce $\hat B \Pi(q^2)^{{\rm (inst)}}$ to the right hand side of Eq.\eqref{eq:newsr} directly
as in Ref.\cite{Shuryak:1982qx,Dorokhov:1989zw,Forkel:1993hj}, and we still use spike distribution.
All Borel-transformed instanton contributions are listed as follows
\begin{gather}
\label{eq:inst1}
\hat B \Pi(q^2)^{\textrm{Fig.\ref{fig:inst}(a)}} =\frac{4 \bar n}{m_q^{\star 2}\bar\rho^6} \me^{-\xi} \xi^5
\left[\xi(1+4\xi) K_0(\xi)+(2+3\xi+4\xi^2)K_1(\xi)\right],\\
\label{eq:inst2}
\hat B \Pi(q^2)^{\textrm{Fig.\ref{fig:inst}(b)}}=\frac{\bar n \pi}{m_q^{\star 2} \bar \rho^2}\langle \alpha_s G^2\rangle \,\me^{-\xi}  \xi^3 \left[ K_0(\xi)+K_1(\xi)\right],\\
\hat B \Pi(q^2)^{\textrm{Fig.\ref{fig:inst}(c)}}=\frac{2048}{25\pi} \frac{\bar n}{\bar \rho^4}\frac{m_q}{m^\star_q} \me^{-\xi}\xi^{3}(1+4\xi)K_{1/2}(\xi),\\
\hat B \Pi(q^2)^{\textrm{Fig.\ref{fig:inst}(d)}}=-64 \bar n \pi^2 \langle m_q \bar qq\rangle \,\me^{-\xi}\xi K_0(\xi),\\
\hat B \Pi(q^2)^{\textrm{Fig.\ref{fig:inst}(e)}}=-\frac{128}{3} \bar n \pi^4 \bar \rho^2 \langle \bar qq\rangle^2 \,\me^{-\xi}\xi K_0(\xi),
\end{gather}
where $\xi=\bar\rho^2/(2\tau)$.

Finally, we should include instanton contributions in our sum rules very carefully  to avoid double counting. If we accept
the assumption that  operator  condensates  are induced by  instantons, then the contributions from Fig.\ref{fig:inst}(a) and (b) (Fig.\ref{fig:inst}(c) and (d))
have double counting, and we should only consider the former contribution. Fig.\ref{fig:inst}(e) is a special case. In the OPE, the four quark condensate can be composed of the same four quarks. Such a four quark condensate cannot be induced in the single instanton picture because  one quark can occupy only one zero-mode state in an instanton,  thus it seems that we should still retain
this Fig.\ref{fig:inst}(e) contribution. 
However, the quark effective mass takes  the collective contribution of all instantons other than the leading one in SIA, thus we cannot
exclude the possibility of double counting between Fig.\ref{fig:inst}(a) and (e),  an issue that deserves further study beyond the scope of this paper. Luckily,
in the present case, including/excluding the contribution from Fig.\ref{fig:inst}(e) does not significantly influence
sum rules in most cases (e.g. $(\bar uug+\bar ddg)/\sqrt 2$), thus we will include Fig.\ref{fig:inst}(e) in our calculation.

Based on the above  considerations, we take the following final instanton contributions for current $j_1$:
\begin{equation}
\hat B \Pi(q^2)^{\textrm{(inst)}}_{j_1}=\hat B \Pi(q^2)^{\textrm{Fig.\ref{fig:inst}(a)}}
+\hat B \Pi(q^2)^{\textrm{Fig.\ref{fig:inst}(c)}}
+\hat B \Pi(q^2)^{\textrm{Fig.\ref{fig:inst}(e)}},
\end{equation}
where all $q=u$, while for $j_2$
\begin{equation}
\hat B \Pi(q^2)^{\textrm{(inst)}}_{j_2}=\hat B \Pi(q^2)^{\textrm{Fig.\ref{fig:inst}(c)}}
+\hat B \Pi(q^2)^{\textrm{Fig.\ref{fig:inst}(e)}},
\end{equation}
where all $q=s$.

Now the mass of the hybrid $m_H$ can be expressed in the form of
\begin{equation}
\label{eq:sr}
m_H=\sqrt{\frac{R_{1}(\tau,s_0)^{\textrm{(OPE)}}+(-\partial/\partial \tau)\hat B \Pi(q^2)^{\textrm{(inst)}}}{R_0(\tau,s_0)
^{\textrm{(OPE)}}+\hat B \Pi(q^2)^{\textrm{(inst)}}}},
\end{equation}
where the moments $R_k^{\textrm{(OPE)}}$ is defined as
\begin{equation}
\label{eq:rk}
R_k(\tau, s_0)^{\textrm{(OPE)}} =\frac{1}{\pi} \int_0^{s_0} \dif s\, s^k \me^{-s\tau} {\rm Im}\Pi(s)^{\textrm{(OPE)}}, ~k=0,1.
\end{equation}

The single narrow resonance spectral density \eqref{eq:single} is an over-simplified model, so sometimes a Breit-Wigner (BW)  form spectral
density is used in QCD sum rules\cite{Dominguez:1984eh,Iqbal:1995gp,Iqbal:1995qc,Elias:1997ya,Leupold:1997dg,Kampfer:2003sq}.
In this paper, we also  study whether the hybrid mass prediction is sensitive to this form of the spectral
density. The only thing we need to do is replace $\delta(s-m_H^2)$ in the spectral density with
\begin{equation}
\frac{1}{\pi}\frac{m_H\Gamma}{(s-m_H^2)^2+m^2_H \Gamma^2},
\end{equation}
where $\Gamma$ is the width of the hybrid\cite{Erkol:2008gp}. Notice that if $\Gamma\to 0$, we can get a delta-type function again.

In this BW-type spectral density model, we should compare
\begin{equation}
\label{eq:bw}
\frac{\int_0^{s_0} s\,\me^{-s\tau}/[(s-m_H^2)^2+m^2_H \Gamma^2] \dif s}{\int_0^{s_0} \me^{-s\tau}/[(s-m_H^2)^2+m^2_H \Gamma^2] \dif s}
\end{equation}
with the square of the right hand side of Eq.\eqref{eq:sr} to get the QCD sum rules. In the QCD sum rules window, the two
quantities should be equal.

Before presenting the numerical results, we should fix the values of various phenomenological parameters
which appear in our sum rules. The condensates and constants
 are chosen as follows\cite{Shifman:1978bx,Shifman:1978by}
\begin{equation}
\begin{split}
&\Lambda_{{\rm QCD}}=0.2\,{\rm GeV},~~m_u=5\,{\rm MeV},~~
m_s=150~{\rm MeV},\\
&\langle \bar uu\rangle =-(0.25\,{\rm GeV})^3,~~\langle \bar ss\rangle=0.8\langle \bar
uu\rangle, ~~\langle \alpha_s G^2\rangle/\pi=0.012\,{\rm GeV}^4.
\end{split}
\end{equation}

Renormalization-group (RG) improvement  of the sum rules amounts to substitutions $\mu^2\to1/\tau$ in Eq.\eqref{eq:impi0}, i.e.,
\begin{equation}
\alpha_s(\mu^2)\rightarrow \alpha_s(1/\tau)=-\frac{4\pi}{9\ln(\tau\Lambda_{{\rm QCD}}^2)}.
\end{equation}

\begin{figure}[htbp]
\centering
\includegraphics[scale=0.9]{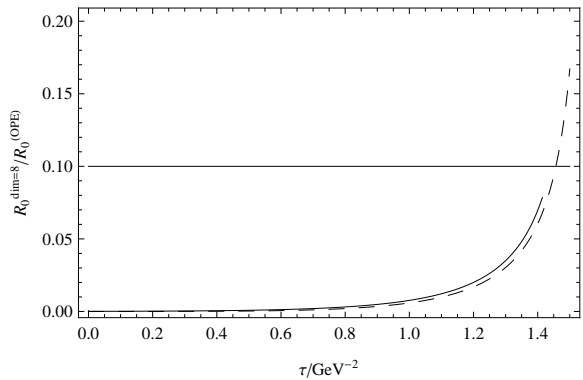}
\caption{\label{fig:uwin} The solid line, the dashed line denote $R_0^{\textrm{dim=8 operator}}/R_0^{{\rm (OPE)}}$ for $j_2$ with $s_0=6.0\,{\rm GeV}^2$
and $s_0=12.0\,{\rm GeV}^2$ respectively.
}
\end{figure}

\begin{figure}[htbp]
\centering
\includegraphics[scale=0.9]{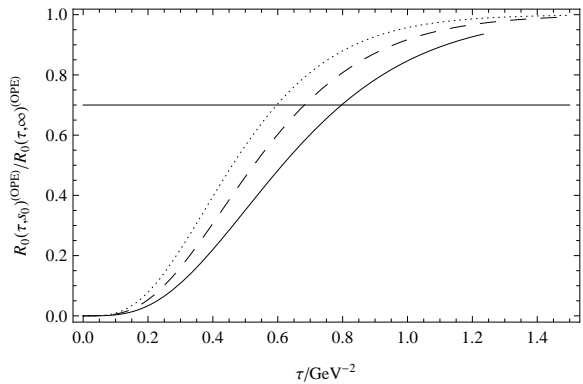}
\caption{\label{fig:lwin} The solid line, the dashed line and the dotted line denote $R_0(\tau,s_0)^{{\rm (OPE)}}/R_0(\tau,\infty)^{{\rm (OPE)}}$ for $j_1$ with $s_0=6.0\,{\rm GeV}^2$, 7.0\,GeV$^2$ and $8.0\,{\rm GeV}^2$ respectively. }
\end{figure}

Finally, we should determine the QCD sum rules window. To establish our sum rules, the Borel parameter $\tau$ should not be too large,
else the convergence of $R_k(\tau,s_0)^{\textrm{(OPE)}}$ will be destroyed because of  effects from the omitted higher dimension condensates in our calculation.
Luckily, in the present cases, the convergence of the OPE series is very good because of the small value of light quark masses.
We make an estimate by demanding that the contribution from the dimension 8 operator, i.e., $\langle m_q \bar qq\rangle^2$, for the sum-rule moment $R_0$ should
be less than 10\% of the total OPE contribution. From Fig.\ref{fig:uwin}, we find $\tau=1.4\,{\rm GeV}^{-2}$ is an appropriate upper bound for $j_2$
(for current $j_1$, we can choose a larger value for the upper bound.).
Meanwhile, $\tau$ should not be too small, else the continuum contribution will
be too large. We demand the continuum contribution is less than 30\% of the total contributions\cite{Ball:2004rg,Zhang:2009qb}.
From Fig.\ref{fig:lwin} we find  $\tau=0.8\,{\rm GeV}^{-2}$ is an appropriate lower bound for $j_1$ with $s_0=6.0\,{\rm GeV}^2$.
If we choose a larger $s_0$, the lower bound of $\tau$ can be reduced to a smaller value. For the current $j_2$, the lower bound of $\tau$
is almost the same. Finally, we also should remember the continuum threshold $s_0$ should be larger than $m^2_H$.

\begin{figure}[htbp]
\centering
\includegraphics[scale=0.9]{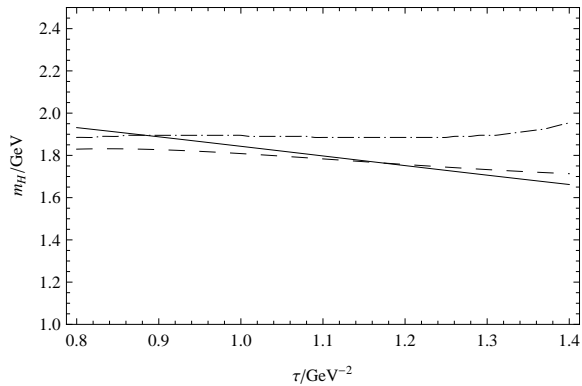}
\caption{\label{fig:1} The sum rules for $j_1$. The solid line, the dashed line and the dot-dashed line denote the sum rule without instanton
contributions, with instanton contributions and the BW-type sum rule (with instanton contributions) respectively. We choose $s_0
=6.0\,\textrm{GeV}^2$ and $\Gamma=0.04\,\textrm{GeV}$.}
\end{figure}

\begin{figure}[htbp]
\centering
\includegraphics[scale=0.9]{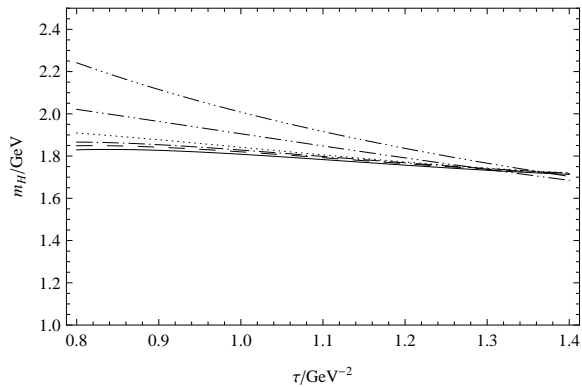}
\caption{\label{fig:2} The sum rules with instanton contributions for $j_1$ with $s_0=\{6.0,~7.0,~8.0,~\infty\}\,{\rm GeV}^2$
for the \{solid line, dashed line, dot-dashed line, dotted line\} respectively, and the sum rules without instanton contributions
for $j_1$ with $s_0=\{7.0,~\infty\}\,{\rm GeV}^2$ for the \{dot-dot-dashed, dot-dot-dot-dashed line\} respectively.}
\end{figure}

In Fig.\ref{fig:1} we show the sum rules for $j_1$. From this figure we find that without instanton contributions, the
sum rule is very unstable. After including instanton contributions, we find an improved sum rule which gives a mass at about 1.83\,GeV
for the scalar hybrid $(\bar uug+\bar ddg)/\sqrt 2$. We should emphasize that, after including instanton contributions, the
sum rule is less sensitive to variation in $s_0$ than the sum rule without these effects (see Fig.\ref{fig:2}).
We also show the BW-type sum rule in Fig.\ref{fig:1}. The dot-dashed lines
shows the fitted mass for the hybrid in demanding that the difference between Eq.\eqref{eq:bw} and the square of the
right hand side of Eq.\eqref{eq:sr} is less than 0.01\,GeV$^2$. We find a heavier mass for the hybrid, about 1.88\,GeV.
This tendency of width effects to increase the sum-rule mass determination has also been observed for $0^{++}$  $\bar q q$ mesons \cite{Elias:1998bq}.
We also find for larger or smaller $\Gamma$, the sum rules lose their stability in the sum rules window.

\begin{figure}[htbp]
\centering
\includegraphics[scale=0.9]{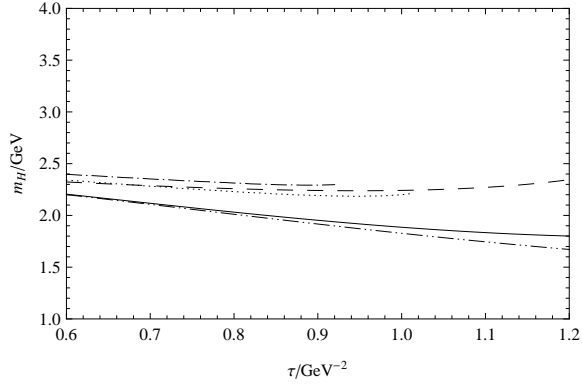}
\caption{\label{fig:fig3} The sum rules for $j_2$. The solid line, the dashed line denote sum rule with and without instanton
contributions  ($s_0=8.0\,\textrm{GeV}^2$) respectively. The dotted line denotes the BW-type sum rule ($s_0=8.0\,\textrm{GeV}^2$,
$\Gamma=0.08\,{\rm GeV}$).
The dot-dot-dashed line and the dot-dashed line denote sum rule ($s_0=8.0\,\textrm{GeV}^2$) and BW-type sum rule
($s_0=8.0\,\textrm{GeV}^2, \Gamma=0.11\,\textrm{GeV}^2$) with instanton contributions
(without contribution from Fig.\ref{fig:inst}(e)) respectively.
}
\end{figure}

From Fig.\ref{fig:fig3} we find after including the instanton contributions, the stability of the mass sum rule of $j_2$ becomes
questionable. This fact may suggest that there are still other effects which are missing in our sum rules, a point which needs further study.
However, if we choose $\Gamma=0.08\,{\rm GeV}$, we find the result for the BW-type sum rule in the
region $0.6\,{\rm GeV^2}<\tau <1\,{\rm GeV^2}$ is almost the same as the mass sum rule without instanton effects in Ref.\cite{Huang:1998zj},
which gives a mass at about 2.20-2.30\,GeV.

If we omit the contribution from Fig.\ref{fig:inst}(e), the sum rule becomes more unstable, but the BW-type sum rule in the region
$0.6\,{\rm GeV^2}<\tau< 0.9\,{\rm GeV}^2$ gives a mass of about 2.30-2.35\,GeV.

\section{The Flavor Octet and Singlet of the Scalar Hybrid}

In the previous section, studying the sum rules for current $j_1$ and $j_2$ revealed that sum rules support the existence of the state associated
with $j_1$  while the one associated with $j_2$ needs further confirmation.

The current $j_1$ and $j_2$ have the same quantum numbers, so they will mix with each other.
In this section, we consider this mixing effect to predict the mass of hybrid in flavor octet and flavor singlet configurations.
For this purpose, we consider the  isospin 0 scalar hybrid current in the general form
\begin{equation}
j(x)=\frac{g}{\sqrt{2c_1^2+c^2_2}}\left\{ c_1\left[\bar u (x) \sigma_{\mu\nu}G^a_{\nu\mu}(x)T^a
u(x)+\bar d (x) \sigma_{\mu\nu}G^a_{\nu\mu}(x)T^a d(x)\right]+c_2 \bar s(x)\sigma_{\mu\nu}G^a_{\nu\mu}(x)T^a s(x)\right\},
\end{equation}
where $c_1$ and $c_2$ are two adjustable parameters. For $c_1=1$ and $c_2=0$ ($c_1=0$ and $c_2=1$), we get our current
$j_1$ ($j_2$) studied in the previous section. For $c_1=1$ and $c_2=-2$ we get the flavor octet hybrid, and for $c_1=c_2=1$,
the hybrid becomes a flavor singlet one.

The mixing effect between $j_1$ and $j_2$ can be described by the correlator
\begin{equation}
\Pi(q^2)^{{\rm (mix)}}=\mi\int\dif^4 x\,\me^{\mi q x}\langle 0|Tj_1(x)j_2^\dagger(0)|0\rangle,
\end{equation}
which will not receive any contribution from perturbative theory. But it does receive instanton contributions, which
is similar to contributions from Fig.\ref{fig:inst}(a), with one $s$-quark loop and one $u$-quark (or $d$-quark) loop in
zero-mode states. The contribution can be read from Eq.\eqref{eq:inst1}
\begin{equation}
\hat B\Pi(q^2)^{\textrm{(inst)}}_{\textrm{mix}}=\hat B\Pi(q^2)^{\textrm{Fig.\ref{fig:inst}(a)}}
\end{equation}
with a replacement rule: $m_q^{\star 2}\to m_u^\star m_s^\star$.

Combining all contributions together, we get the final result for current $j$:
\begin{equation}
\begin{split}
{\rm Im}\Pi(s)^{\textrm{(OPE)}}&= \frac{2 c^2_1}{2 c_1^2+c^2_2} \left[\frac{\alpha_s}{24\pi^2}s^3 + 4 \alpha_s m_u\langle \bar uu\rangle s
-m^2_u\langle \alpha_s G^2\rangle +\frac{8\alpha_s\pi^2}{3} m^2_u \langle \bar uu\rangle^2 \delta(s)\right]\\
&+\frac{c^2_2}{2 c_1^2+c^2_2} \left[\frac{\alpha_s}{24\pi^2}s^3 + 4 \alpha_s m_s\langle \bar ss\rangle s
-m^2_s\langle \alpha_s G^2\rangle +\frac{8\alpha_s\pi^2}{3} m^2_s \langle \bar ss\rangle^2 \delta(s)\right],
\end{split}
\end{equation}
and
\begin{equation}
\hat B\Pi(q^2)^{\textrm{(inst)}}_j=\frac{2 c^2_1}{2 c_1^2+c^2_2} \hat B\Pi(q^2)^{\textrm{(inst)}}_{j_1}
+\frac{c^2_2}{2 c_1^2+c^2_2} \hat B\Pi(q^2)^{\textrm{(inst)}}_{j_2}
+ \frac{4 c_1 c_2}{2 c_1^2+c_2^2}\hat B\Pi(q^2)^{\textrm{(inst)}}_{\textrm{mix}}.
\end{equation}

\begin{figure}[htbp]
\centering
\includegraphics[scale=0.9]{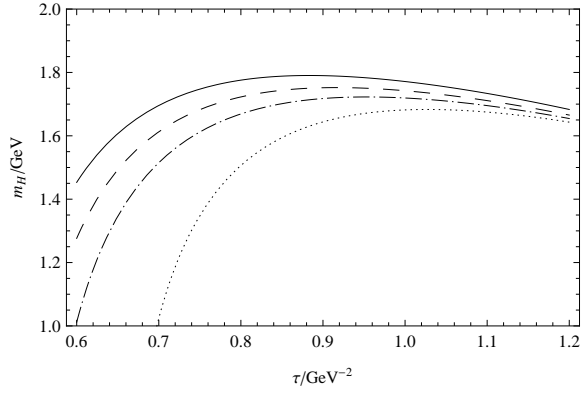}
\caption{\label{fig:4} The sum rules for the flavor octet hybrid
with $s_0=\{6.0,~7.0,~8.0,~\infty\}\,{\rm GeV}^2$
for the \{solid line, dashed line, dot-dashed line, dotted line\} respectively.}
\end{figure}

\begin{figure}[htbp]
\centering
\includegraphics[scale=0.9]{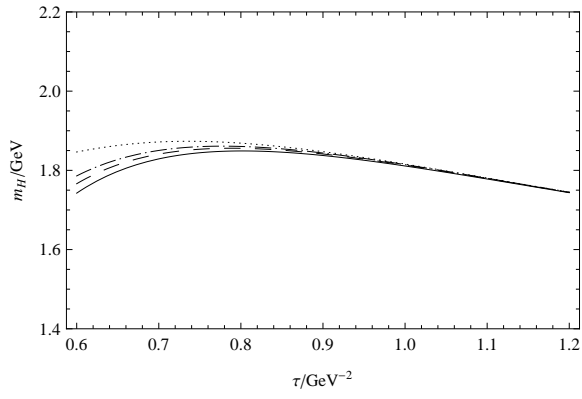}
\caption{\label{fig:5} The sum rules for the flavor singlet hybrid
with $s_0=\{9.0,~10.0,~11.0,~\infty\}\,{\rm GeV}^2$
for the \{solid line, dashed line, dot-dashed line, dotted line\} respectively.}
\end{figure}

We plot the sum rules for the flavor octet isoscalar hybrid in Fig.\ref{fig:4}
and the flavor singlet state in Fig.\ref{fig:5}. From Fig.\ref{fig:4} we find
the mass upper bound ($s_0\to\infty$) of the flavor octet hybrid is smaller than that with a
finite $s_0$. Since this behaviour is unphysical, we conclude that this state is not
supported by QCD sum rules.

\begin{figure}[htbp]
\centering
\includegraphics[scale=0.9]{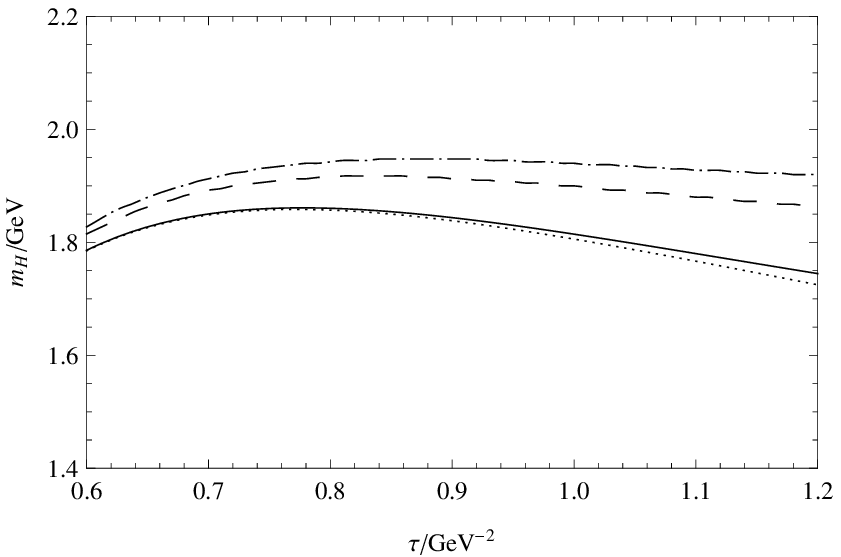}
\caption{\label{fig:6} The sum rules for the flavor singlet hybrid.
The solid line, dashed line denote sum rule ($s_0=11.0\,{\rm GeV}^2$), BW-type sum rule ($s_0=11.0\,{\rm GeV}^2$ and $\Gamma=0.04\,{\rm GeV}$)
respectively. The dotted line and dot-dashed line denote sum rule ($s_0=11.0\,{\rm GeV}^2$),
BW-type sum rule ($s_0=11.0\,{\rm GeV}^2$ and $\Gamma=0.06\,{\rm GeV}$) without contribution from Fig.\ref{fig:inst}(e) respectively.
}
\end{figure}

From Fig.\ref{fig:5} we find the mass of the flavor singlet hybrid is not very sensitive to $s_0$ in our
sum rules window. Choosing $s_0=11.0\,{\rm GeV}^2$ as a typical $s_0$ value, from Fig.\ref{fig:6} we get
a flavor singlet hybrid mass at 1.86\,GeV. If we use a BW-type sum rule with $\Gamma=0.04\,{\rm GeV}$, we
get a heavier mass, about 1.92\,GeV.

Omitting the contribution from Fig.\ref{fig:inst}(e) does not change the sum rule a lot, but it does allow
a large width ($\Gamma =0.06\,{\rm GeV}$), which leads to a mass at about 1.95\,GeV.

\section{Mixing with Scalar Glueball}

The hybrid currents $j_1$ and $j_2$ may also mix with the glueball current (\ref{glueball}). From the perspective of perturbation theory, such a mixing is chirally suppressed, so the mixing must be non-perturbative if it is not small. In our framework, the mixing between the scalar hybrid and the scalar glueball may occur through instanton and quark condensate effects. In this section, we discuss
how instanton  and quark condensate contributions affect the sum rules. With this motivation, we consider a scalar current\cite{Kisslinger:2001pk}

\begin{equation}\label{glueball}
j_{0^{++}}=\beta j_{\textrm{hybrid}}+(1-|\beta|) M_0 j_{\textrm{glueball}},
\end{equation}
where $j_{\textrm{glueball}}=\alpha_s G^2$ is the scalar glueball current, $M_0$ is a parameter which has the dimension of the mass while
$\beta$ is a parameter which can run from -1 to 1.

All contributions for the glueball correlator already exist in literature, e.g., the OPE contribution to the
glueball current correlator reads (up to dimension 8 condensate) \cite{Harnett:2000xz,Harnett:2008cw}:
\begin{equation}
\begin{split}
R_0(\tau,s_0)_{\rm{glueball}}^{\textrm{(OPE)}}=&\int_{0}^{s_{0}}\dif s\,s^{2}\me^{-s\tau}\left\{2\left(\frac{\alpha_{s}}{\pi}\right)^{2}
\left[1+\frac{659}{36}\frac{\alpha_{s}}{\pi}+247.48\left(\frac{\alpha_{s}}{\pi}\right)^{2}\right]
-4\left(\frac{\alpha_{s}}{\pi}\right)^{3}\left(\frac{9}{4}+65.781\frac{\alpha_{s}}{\pi}\right)\ln\frac{s}{\mu^{2}}\right.\\
&-10.125\left.\left(\frac{\alpha_{s}}{\pi}\right)^{4}\left(\pi^{2}-3\ln^{2}\frac{s}{\mu^{2}}\right)\right\}
+9\pi\left(\frac{\alpha_{s}}{\pi}\right)^{2}\langle\alpha_{s}G^{2}\rangle\int_{0}^{s_{0}}\dif s\,\me^{-s\tau}
+8\pi^{2}\left(\frac{\alpha_{s}}{\pi}\right)^{2}\langle\mathcal{O}_{6}\rangle+8\pi^{2}
\frac{\alpha_{s}}{\pi}\langle\mathcal{O}_{8}\rangle\tau,
\end{split}
\end{equation}
where $\langle\mathcal {O}_{6}\rangle=\langle g_{s}f_{abc}G^{a}_
{\mu\nu}G^{b}_{\nu\rho}G^{c}_{\rho\mu}\rangle=(0.27\,{\rm{GeV^{2}}})\langle\alpha_{s}G^{2}\rangle$ and $\langle\mathcal {O}_{8}\rangle=
14\langle(\alpha_{s}f_{abc}G^{a}_{\mu\nu}G^{b}_{\nu\rho})^{2}\rangle
-\langle(\alpha_{s}f_{abc}G^{a}_{\mu\nu}G^{b}_{\rho\lambda})^{2}\rangle=\frac{9}{16}(\langle\alpha_{s}G^{2}\rangle)^{2}$,
and the instanton contribution to the glueball current correlator is\cite{Forkel:2000fd}:
\begin{equation}
\hat B\Pi(q^2)_{\textrm{glueball}}=\frac{256 \bar n\pi^2}{\bar \rho^2} \me^{-\xi} \xi^5\left[K_0(\xi)+\left(1+\frac{1}{2\xi}\right)
K_1(\xi)\right].
\end{equation}

The only unknown new term is the mixing contribution between hybrid and glueball
\begin{equation}
\Pi(q^2)_{\textrm{H-GB}}=
\mi \int\dif^4 x\,\me^ {\mi
q x} \langle0| T g\bar q(x) \sigma_{\mu\nu} G^a_{\nu\mu}(x) T^a q(x) \alpha_s G(0)^2|0\rangle,
\end{equation}
after some calculations, we get its contribution:
\begin{equation}
\hat B\Pi_{\textrm{H-GB}}^{\textrm{(inst)}}+R_0(\tau,s_0)^{\textrm{(OPE)}}_{\textrm{H-GB}}=-\frac{16\bar n\pi}{m^\star_q\bar\rho^4}\{\xi^3-\xi^4
\me^{-\xi}[\xi(1+4\xi)K_0(\xi)+(2+3\xi+4\xi^2)K_1(\xi)]
\}
-8\pi\left(\frac{\alpha_s}{\pi}\right)^2\frac{\langle \bar q q\rangle}{\tau^2}[1-\me^{-s_0 \tau}(1+s_0\tau)]
.
\end{equation}
Although the mixed condensate can also avoid chiral suppression, its leading order contribution is zero after Borel transforming.

Before proceeding with the numerical calculations, let us fix the parameter $M_0$. Actually, the value of
$M_0$ does not affect the sum rules because there is another parameter $\beta$.  For convenience (so that $\beta$ can approximately represent the mixing intensity), we choose  $M_0=0.02\,\textrm{GeV}$ so that the leading contributions of hybrid correlator and glueball correlator
($\alpha_s s^3/(24\pi^2)$ and $M_0\cdot 2\alpha_s^2 s^2/\pi$) are
the same order of magnitude at this scale, where we choose $\mu=1\,{\textrm{GeV}}$ in this section, and  $\alpha_s$ is fixed to 0.517.

\begin{figure}[htbp]
\centering
\includegraphics[scale=0.9]{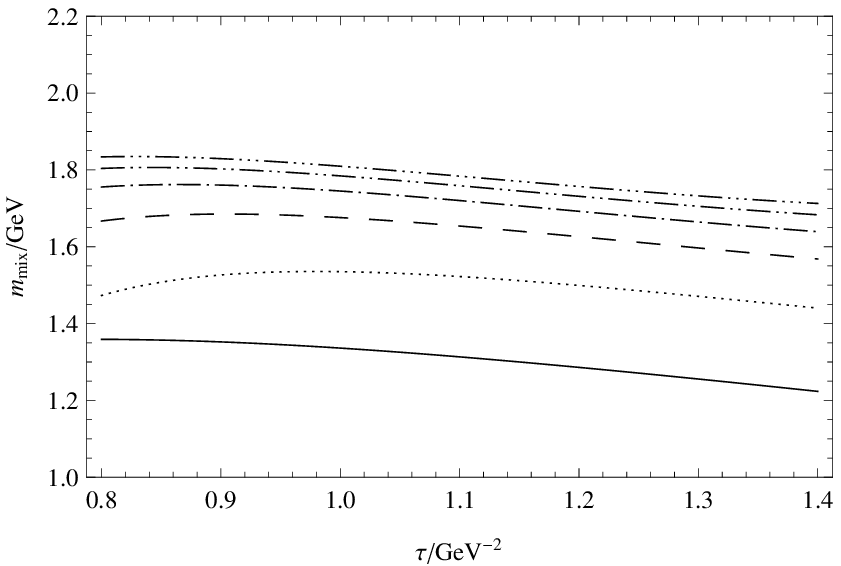}
\caption{\label{fig:7} The sum rules for the mixing state of $\frac{1}{\sqrt 2}(\bar uug+\bar ddg)$ and
the scalar glueball ($j=\beta j_1+(1-|\beta|)j_{\text{glueball}}$, $s_0=6.0\,\textrm{GeV}^2$) with $\beta=\{
0,~0.2,~0.4,~0.6,~0.8,~1\}$ for the \{solid line,
dotted line, dashed line, dot-dashed line, dot-dot-dashed line, dot-dot-dot-dashed line\} respectively. }
\end{figure}

\begin{figure}[htbp]
\centering
\includegraphics[scale=0.9]{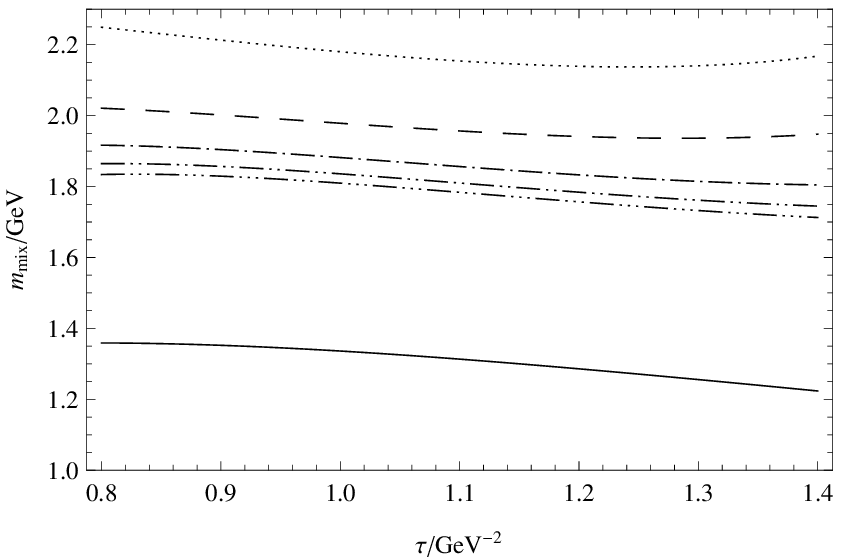}
\caption{\label{fig:8} The sum rules for the mixing state of $\frac{1}{\sqrt 2}(\bar uug+\bar ddg)$ and
the scalar glueball ($j=\beta j_1+(1-|\beta|)j_{\text{glueball}}$, $s_0=6.0\,\textrm{GeV}^2$) with $\beta=\{
0,~-0.2,~-0.4,~-0.6,~-0.8,~-1\}$ for the \{solid line,
dotted line, dashed line, dot-dashed line, dot-dot-dashed line, dot-dot-dot-dashed line\} respectively. }
\end{figure}

In Fig.\ref{fig:7} and Fig.\ref{fig:8} we show our result for the mixed state of $\frac{1}{\sqrt 2}(\bar uug+\bar ddg)$
and glueball. From Fig.\ref{fig:7} we find that if $0<\beta<1$, then the mass of the mixed state
will lie in the region 1.35-1.83\,GeV, which covers the masses of $f_0(1370)$, $f_0(1500)$ and $f_0(1710)$.
For example, the mass of the mixed state with $\beta=0.4$ is very close to the mass of $f_0(1710)$.
From Fig.\ref{fig:8} we find if $0>\beta>-1$, then the mass of the mixed state will lie in the region
1.83-2.2\,GeV. Actually we find for $\beta>-0.2$, the sum rules are unstable. This fact is quite understandable. 
In the region $0>\beta>-1$, when $\beta$ goes to zero, the content of the glueball increases and the expected mass goes up. Meanwhile £¬according to the result in  the region $0<\beta<1$,  the expected mass must go down when $\beta$ is very close to zero,. This contradictory causes the sum rules unstable. 
 In other words, the content of the hybrid must dominate in the region $0>\beta>-1$ .  Clearly, after the mixing is taken into account, there are two states.
One, dominated by the glueball content, is lighter, and may be  $f_0(1370)$, $f_0(1500)$ and $f_0(1710)$; another one, dominated by hybrid content, is in region 1.83-2.2\,GeV.
It seems straightforward to extend our discussion to the mixing between the scalar hybrid and the normal $\bar q q$ scalar, however, there is a problem for such an extension because in our framework it is hard to identify the  $\bar q q$ scalar in the group of $f_0(1370)$, $f_0(1500)$ and $f_0(1710)$ and in the region $2000\,{\rm MeV}\sim 2300\,{\rm MeV}$ (both of which may mix with the hybrid). Certainly, this argument is also valid
for the mixing between the glueball and the hybrid if there are two glueballs in these two groups respectively.

\begin{figure}[htbp]
\centering
\includegraphics[scale=0.9]{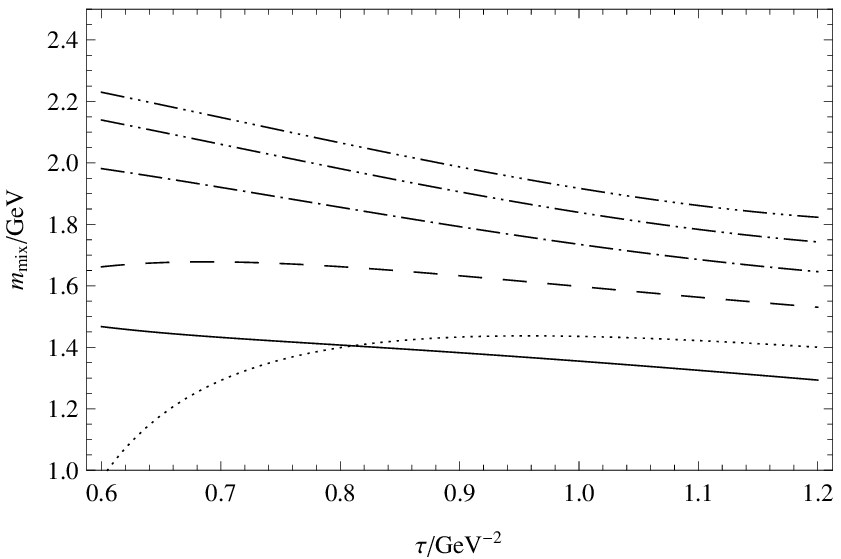}
\caption{\label{fig:9} The sum rules for the mixing state of $\bar ssg$ and
the scalar glueball ($j=\beta j_2+(1-|\beta|)j_{\text{glueball}}$, $s_0=8.0\,\textrm{GeV}^2$) with $\beta=\{
0,~0.2,~0.4,~0.6,~0.8,~1\}$ for the \{solid line,
dotted line, dashed line, dot-dashed line, dot-dot-dashed line, dot-dot-dot-dashed line\} respectively. }
\end{figure}

\begin{figure}[htbp]
\centering
\includegraphics[scale=0.9]{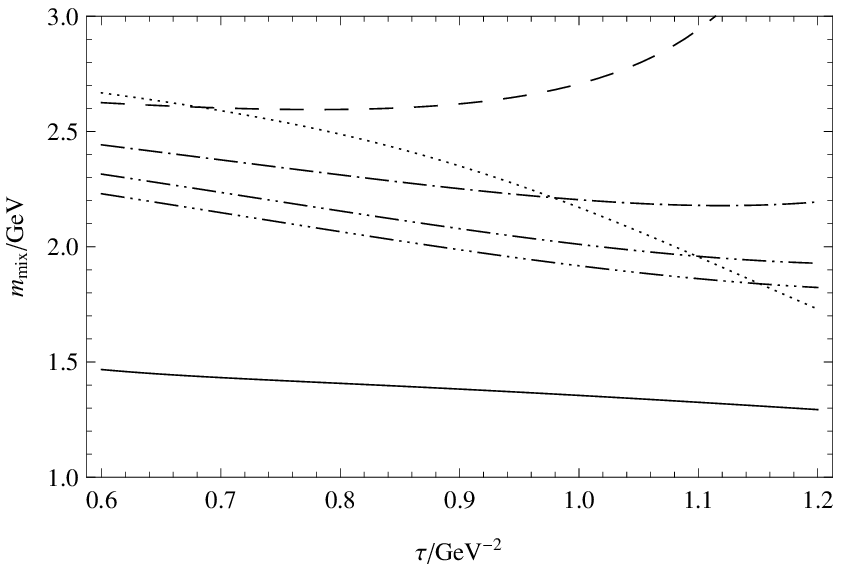}
\caption{\label{fig:10} The sum rules for the mixing state of $\bar ssg$ and
the scalar glueball ($j=\beta j_2+(1-|\beta|)j_{\text{glueball}}$, $s_0=8.0\,\textrm{GeV}^2$) with $\beta=\{
0,~-0.2,~-0.4,~-0.6,~-0.8,~-1\}$ for the \{solid line,
dotted line, dashed line, dot-dashed line, dot-dot-dashed line, dot-dot-dot-dashed line\} respectively. }
\end{figure}

Similarly, from Fig.\ref{fig:9} and Fig.\ref{fig:10} we learn that the mass of the mixed state of $\bar ssg$
and glueball is in the range 1.4-2.6\,GeV, which may also shed light on the explanation of the contents of these light
$f_0$ mesons.

\section{Discussion and Summary}
In this paper, we have calculated instanton contributions to the sum rules for the scalar hybrid.
We find that instanton effects play an important role in the sum rule mass predictions for $(\bar uug+\bar ddg)/\sqrt 2$
hybrids. The sum rules for the mass of the scalar hybrid  become quite stable and predictable
 after instanton contributions are included. The mass of the $(\bar uug+\bar ddg)/\sqrt 2$ hybrid  is around  1.83\,GeV.
 However, the instanton effects for the sum rules of the $\bar ssg$ hybrid are not stable.
 Although our analysis seems to suggest a heavier mass for the $\bar ssg$ scalar hybrid ($m_H\geq 2\,{\rm GeV}$), it still needs further study.

We also considered non-zero width effects for the sum rules. We find that the $(\bar uug+\bar ddg)/\sqrt 2$
hybrid can be compatible with a very small width ($\Gamma=0.04\,{\rm GeV}$). After considering this effect,
the predicted mass increases about 0.05\,GeV, and the sum rules become more stable. Larger or smaller width $\Gamma$ will
make the sum rule unstable, but we cannot exclude such a possibility. For the $\bar ssg$ scalar hybrid, instanton
effects decrease the stability of sum rule a little, but if
we consider an appropriate width, the sum rule with instantons can be very close
to the sum rule without instantons in the  narrow resonance ansatz, which gives a mass at about 2.20-2.30\,GeV.

Instantons also can induce the mixing between $j_1$ and $j_2$, which  leads
to a flavor singlet hybrid with mass around 1.9\,GeV,  while the flavor octet hybrid is not supported
by QCD sum rules.

Finally, we studied the mixing effects between scalar hybrid and glueball currents. The mixing between the $(\bar uug+\bar ddg)/\sqrt{2}$($\bar ssg$) and the glueball causes two states, one in the region 1.4-1.8\,GeV(1.4-2.2\,GeV), and the other  in the range  1.8-2.2\,GeV(2.2-2.6\,GeV).

Understanding the crowded $0^{++}$ meson spectrum in the regions 1.3-1.7\,GeV and 2.0-2.3\,GeV is an important task in particle physics, both in the naive quark model and in non-$\bar qq$ meson model. Our result may give possible interpretations
of some of these mesons. For example, $f_0(2020)$ is a particle listed in
the latest version of Review of Particle Physics that still requires confirmation\cite{Nakamura:2010zzi}.
Its mass is 1.99\,GeV, and width is 0.55\,GeV. Notice that our result shows that the flavor singlet scalar
hybrid mass is very close to the mass of $f_0(2020)$ with the same quantum numbers. This may lead to a possible
explanation that part of the contents of $f_0(2020)$ is a scalar hybrid. Based on our results, even $f_0(1710)$ ($m=1.72\,{\rm GeV}$, $\Gamma
=0.135\,{\rm GeV}$) may have partial hybrid content.

\begin{acknowledgments}
This work is partly supported by K. C. Wong Magna Fund in Ningbo University and NSFC under grant 11175153/A050202.
\end{acknowledgments}

\end{document}